\documentclass[amsmath,amssymb]{revtex4}

\usepackage{color}

\usepackage{graphicx}

\begin{document}
\title{Optical Density-Enhanced Squeezed Light Generation without Optical Cavities}

\author{You-Lin Chuang,$^{1,2}$ Ray-Kuang Lee,$^{1,2,3,4,}$}\email{rklee@ee.nthu.edu.tw}
\author{Ite A. Yu$^{3,4,}$}\email{yu@phys.nthu.edu.tw}
\affiliation{
$^1$Physics Division, National Center for Theoretical Sciences, Hsinchu 30013, Taiwan\\
$^2$Institute of Photonics Technologies, National Tsing Hua University, Hsinchu 30013, Taiwan\\
$^3$Department of Physics, National Tsing Hua University, Hsinchu 30013, Taiwan\\
$^4$Frontier Research Center on Fundamental and Applied Sciences of Matters, National Tsing Hua University, Hsinchu 30013, Taiwan}

\date{\today}

\begin{abstract}
To achieve high degree of quantum noise squeezing, an optical cavity is often employed to enhance the interaction time between light and matter. Here, we propose to utilize the effect of coherent population trapping (CPT) to directly generate squeezed light without any optical cavity. Combined with the slow propagation speed of light in a CPT medium, a coherent state passing through an atomic ensemble with a high optical density (OD) can evolve into a highly squeezed state even in a single passage. Our study reveals that noise squeezing of more than $10$ dB can be achieved with an OD of $1,000$, which is currently available in experiments. A larger OD can further increase the degree of squeezing. As the light intensity and two-photon detuning are key factors in the CPT interaction, we also demonstrate that the minimum variance at a given OD can be reached for a wide range of these two factors, showing the proposed scheme is flexible and robust. Furthermore, there is no need to consider the phase-matching condition in the CPT scheme. Our introduction of high OD in atomic media not only brings a long light-matter interaction time comparable to optical cavities, but also opens new avenue in the generation of squeezed light for quantum interface.
\end{abstract}

\maketitle

\newcommand{\FigOne}{
\begin{figure}[t]
	\includegraphics[width=8.0cm]{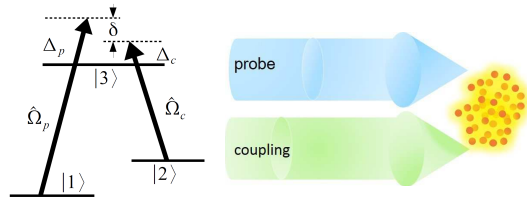}
	\caption{Energy levels and excitations in the CPT system. $\vert 1\rangle$ and $\vert 2\rangle$ are ground states; $\vert 3\rangle$ is an excited state. The probe and coupling fields have the compatible Rabi frequencies of $\Omega_p$ and $\Omega_c$, and the detunings of $\Delta_p$ and $\Delta_c$. They propagate in the same direction and interact with an atomic ensemble.}
\end{figure}}
\newcommand{\FigTwo}{
\begin{figure}[t]
	\includegraphics[width=8.5cm]{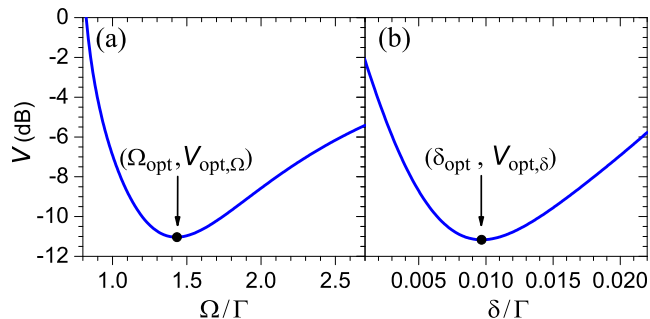}
	\caption{Output variance $V$, defined in Eq.~(\ref{Vopt}), as functions of input Rabi frequency $\Omega$ and two-photon detuning $\delta$. In (a), $\alpha$ (i.e. OD) = 1,000 and $\delta = 0.02\Gamma$; in (b), $\alpha$ = 1,000 and $\Omega = 1.0\Gamma$. The values of minimum variance or maximum squeezing in the two plots are nearly the same. $\Omega_{\rm opt}$ (or $\delta_{\rm opt}$) is the optimum input Rabi frequency (or the optimum two-photon detuning) that minimizes $V$ under a fixed $\delta$ (or $\Omega$).}
\end{figure}}
\newcommand{\FigThree}{
\begin{figure}[t]
	\includegraphics[width=8.5cm]{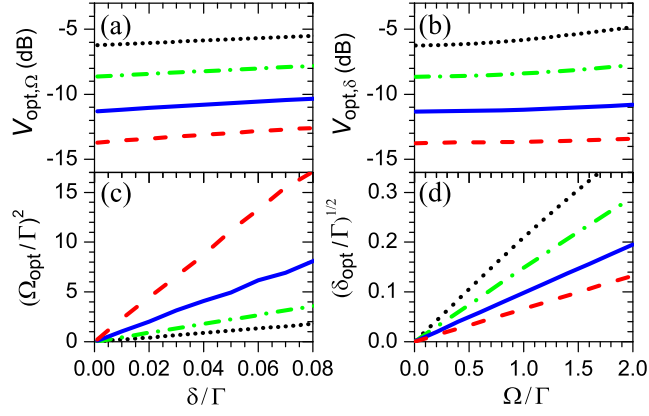}
	\caption{Optimized squeezing at different ODs. Black dotted, green dashed-dotted, blue solid, and red dashed lines represent OD's values of 100, 300, 1,000, and 3,000, respectively. All results were
 obtained by numerically solving Eqs.~(A24)-(A29). (a) The maximum squeezing obtained by scanning all input Rabi frequencies, $V_{{\rm opt},\Omega}$, as a function of two-photon detuning $\delta$; (c) the corresponding $\Omega_{\rm opt}$ versus $\delta$. (b) The maximum squeezing obtained by scanning all two-photon detunings, $V_{{\rm opt},\delta}$, as a function of input Rabi frequency $\Omega$; (d) the corresponding $\delta_{\rm opt}$ versus $\Omega$.}
\end{figure}}
\newcommand{\FigFour}{
\begin{figure}[t]
	\includegraphics[width=8.5cm]{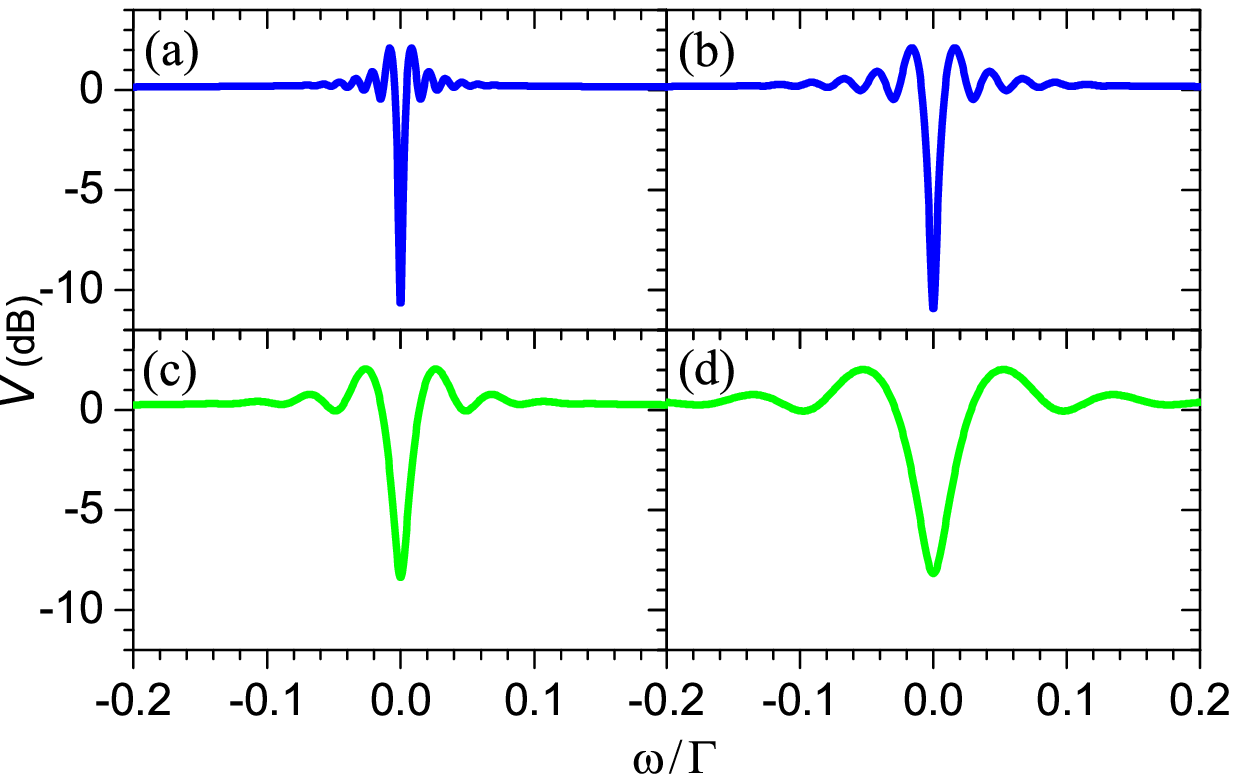}
	\caption{Squeezing spectra of the probe field, i.e. output variance $V$ as functions of noise frequency $\omega$. (a) $\alpha$ (OD) = 1,000, $\Omega$ (input Rabi frequency) = $1.0\Gamma$, and $\delta$ (two-photon detuning) = $0.01\Gamma$; (b) $\alpha$ = 1,000, $\Omega$ = $1.4\Gamma$, and $\delta$ = $0.019\Gamma$; (c) $\alpha$ = 300, $\Omega$ = $1.0\Gamma$, and $\delta$ = $0.019\Gamma$; (d) $\alpha = 300$, $\Omega$ = $1.4\Gamma$, and $\delta$ = $0.043\Gamma$. The four sets of $\Omega$ and $\delta$ all optimize the squeezing at $\omega$ = 0. In each spectrum, the quadrature angle is kept the same.}
\end{figure}}
\section{Introduction}

Even though Heisenberg uncertainty relation sets a fundamental limit on the quantum fluctuations, the noise of light at certain phases can be {\it squeezed} to fall below that of the vacuum state~\cite{squeezedlight}. Generation of squeezed light has provided the platform to test quantum physics from the very beginning~\cite{SQ-1}.  Now as true applications, this non-classical state has also been used to enhance quantum metrology~\cite{metrology, metrology2} and future gravitational wave detection~\cite{gravitation, LIGO}. Quantum noise squeezing has been realized in a variety of physical settings from  optical parametric process~\cite{LIGO, OPA, 15dB, timedelay_sq, modulated_OPO}, four-wave mixing~\cite{Slusher, Lett, FWM}, cavity-QED~\cite{cavityCPT}, soliton propagation~\cite{solitonsqueezing, SITsqueezing}, Bose-Einstein condensate~\cite{BECsqueezing}, and optomechanical system~\cite{optomechanicalsqueezing, optomech2}.

With the process of degenerate parametric down-conversion in a nonlinear crystal placed inside an optical cavity,  optical parametric oscillator (OPO) and optical parametric amplification (OPA) have provided efficient routines to produce high degree of squeezing. In particular, $12.7$ dB squeezing below vacuum fluctuation with a zero-area Sagnac interferometer was implemented, and may lead to advanced gravitational-wave detectors~\cite{LIGO}. Assisted by a doubly resonant, nonmonolithic OPA cavity, up to $15$ dB  squeezing was observed as the state-of-the-art technology~\cite{15dB}. Further enhancement on the degree of squeezing can be achieved with periodically poled nonlinear crystals, via the time-delayed coherent feedback~\cite{timedelay_sq}, or by using periodically modulated driving fields~\cite{modulated_OPO}.

Before being produced with the optical parametric process in a nonlinear crystal~\cite{OPA}, squeezed light was first realized through the four-wave mixing process in an atomic vapor~\cite{Slusher}. Although merely $0.3$ dB squeezing was detected at that time, by considering twin-beam squeezing in the double-$\Lambda$ transition scheme, 8 dB squeezing was achieved with a vapor of rubidium atoms later~\cite{Lett}. In the system of electromagnetically induced transparency (EIT), not only slowing down but also storing and retrieving squeezed-state light pulses have been studied theoretically and experimentally~\cite{EITsqueezing1, EITsqueezing2, squeezingslowlight, diffusionD, sqEIT,Exp_sq1, Exp_sq2, Exp_sq3}. The EIT system plays a unique role as the quantum interface, because of its long-lived atomic ground states associate with the spin coherence~\cite{Quantumstatetransfer, Quantum_Memory}. However, it is not favorable for the direct generation of squeezed light due to its lack of nonlinear interaction between slow light and the medium.

Inspired by the recent experimental advance of high optical density or depth (OD) in atomic ensembles~\cite{OD1, OD2, OD3}, in this work, we study the direct generation of squeezed light under the coherent population trapping (CPT) condition \cite{sqCPT1, sqCPT2}. The CPT system is very similar to the EIT system, formed by the $\Lambda$-type transition scheme as shown in Fig.~1, except that its two optical fields have compatible intensities. Without any optical cavity, we show that a large OD in the CPT system not only results in a very long light-matter interaction time arising from the slow-light effect, but also benefits to the generation of highly-squeezed light induced by the two-photon detuning in the system. This single-passage generation of squeezed light relies on the idea of using a large OD or, equivalently, long medium length to enhance the light-matter interaction time, similar to the concept of using an optical cavity. 

Moreover,  our scheme for squeezed light generation is in analogy to the periodic-poled nonlinear crystal for nonlinear optical processes, but does not require the consideration of phase-matching condition. Compatible to optical parametric processes, an enhancement of more than $10$ dB squeezing is exhibited at the output fields with an OD of $1,000$. Moreover, the obtained squeezing is available for a wide range of input light intensity and two-photon detuning. The squeezing can be reached for a wide range of experimental parameters in our scheme, showing it is flexible and robust. Because OD is scalable with medium length, the best up-to-date squeezing can likely be improved by increasing OD or medium length in the proposed scheme. With such highly-squeezed light generated at the output fields, combined with the inherent capability of storage and retrieval of quantum information carried by light, our work may open a renewed interest in quantum noise reduction, quantum memory, and quantum information manipulation with atomic ensembles.

\FigOne

\section{Quantum theory for coherent population trapping (CPT)}

The CPT system consists of two optical fields interacting with a three-level $\Lambda$-type system as shown in Fig.~1. The two fields, named probe and coupling, drive the transitions of $\vert 1\rangle \rightarrow \vert 3 \rangle$ and  $\vert 2\rangle \rightarrow \vert 3 \rangle$, respectively. Under the rotating-wave approximation, the interaction Hamiltonian of the system is
\begin{eqnarray}
\label{H}
	\hat{H} = -\hbar \left[ \Delta_p \hat{\sigma}_{33}(z,t) 
		+ (\Delta_p - \Delta_c) \hat{\sigma}_{22}(z,t) \right] 
		\hspace*{1.8cm} \nonumber \\
	-{\hbar} \left[ \frac{\hat{\Omega}_p(z,t)}{2} \hat{\sigma}_{31}(z,t)
		+ \frac{\hat{\Omega}_c(z,t)}{2} \hat{\sigma}_{32}(z,t) + H.C. \right]\!\! ,
\end{eqnarray}
where $\Delta_p$ and $ \Delta_c$ are the probe and coupling detunings, $\hat{\sigma}_{ij} \equiv \vert i\rangle\langle j\vert$ ($i, j = 1, 2, 3$) is the atomic operator whose expectation value corresponds to an element of the density-matrix operator, and $\hat{\Omega}_p(z,t)$ and $\hat{\Omega}_c(z,t)$ are the field operators whose expectation values correspond to the probe and coupling Rabi frequencies, respectively. We define $\delta$ ($\equiv \Delta_p - \Delta_c$) as the two-photon detuning.

According to the Hamiltonian in Eq.~(\ref{H}), we can write down the corresponding Heisenberg-Langevin equations for atomic operators as follows.
\begin{eqnarray}
\label{HE1}
	\dfrac{\partial}{\partial t}\hat{\sigma}_{\mu\mu} 
		&=& -\Gamma_{\mu}\hat{\sigma}_{33}
		+ \dfrac{1}{i\hbar}\left[ \hat{\sigma}_{\mu\mu},\hat{H}\right] 
		+ \hat{F}_{\mu\mu}, \\
\label{HE2}
	\dfrac{\partial}{\partial t}\hat{\sigma}_{\mu\nu} 
		&=& -\gamma_{\mu\nu}\hat{\sigma}_{\mu\nu} 
		+ \dfrac{1}{i\hbar}\left[ \hat{\sigma}_{\mu\nu},\hat{H}\right] 
		+ \hat{F}_{\mu\nu} ,~ (\mu\neq \nu)
\end{eqnarray}
where $\gamma_{\mu\nu}$ ($\mu, \nu = 1, 2, 3$) is the relaxation rate of the coherence between states $\vert\mu\rangle$ and $\vert\nu\rangle$, $\Gamma_\mu$ is the decay rate of the population, and $\hat{F}_{\mu\nu}$ is the Langevin noise operator obtained by taking the fluctuation-dissipation theorem into consideration. In this work, we consider $\gamma_{12}$ is negligible~\cite{gamma}. Since $\Gamma$ represents the spontaneous decay rate of the excited state $\vert 3\rangle$, $\Gamma_{3} = \Gamma$ and $\gamma_{23} =\gamma_{13} = \Gamma/2$. The decay rates of $\vert 3\rangle \rightarrow \vert 1\rangle$ and $\vert 3\rangle \rightarrow \vert 2\rangle$ are set the same and, consequently, $-\Gamma_1 = -\Gamma_{2} = \Gamma/2$. The complete equations can be found in Sec.~1 of the Appendix.

The propagations of the probe and coupling fields follow the Maxwell-Schr\"{o}dinger equations given by
\begin{eqnarray}
\label{op}
	&&\left( \dfrac{1}{c}\dfrac{\partial}{\partial t}
		+\dfrac{\partial}{\partial z}\right)\hat{\Omega}_p = 
		i \left( \dfrac{\Gamma \alpha}{2L}\right)  \hat{\sigma}_{13}, \\ 
 \label{oc}
	&&\left( \dfrac{1}{c}\dfrac{\partial}{\partial t}
		+\dfrac{\partial}{\partial z}\right)\hat{\Omega}_c 
		= i \left( \dfrac{\Gamma \alpha}{2L}\right)  \hat{\sigma}_{23},
\end{eqnarray}
where $\alpha$ and $L$ are the OD and length of the medium. For simplicity, we use the same OD in the above two equations under the assumption that the electric dipole moments of probe and coupling transitions are equal.

To calculate the variances of output fields, we apply the mean-field expansion to operators, i.e., each operator $\hat{A}$ is divided into two parts as $\hat{A} = A + \hat{a}$, where $A$ represents the mean-field value and $\hat{a}$ corresponds to the fluctuation operator. Then, one can linearize Eqs.~(\ref{HE1}) and (\ref{HE2}) to arrive at the following equations for the atomic fluctuation operators.
\begin{eqnarray}
\label{s11}
	&&\dfrac{\partial}{\partial t}\hat{s}_{11} = 
		\dfrac{\Gamma}{2}\hat{s}_{33} 
		-\dfrac{i}{2}\Omega_p\hat{s}_{31} 
		+\dfrac{i}{2}\Omega_p^{\ast}\hat{s}_{13} \nonumber \\
	&&~~~~~~~~~~~-\dfrac{i}{2}\sigma_{31}\hat{u}_p 
		+\dfrac{i}{2}\sigma_{13}\hat{u}_p^{\dagger} 
		+\hat{F}_{11}, \\
\label{s22}
	&&\dfrac{\partial}{\partial t}\hat{s}_{22} = 
		\dfrac{\Gamma}{2}\hat{s}_{33} 
		-\dfrac{i}{2}\Omega_c\hat{s}_{32} 
		+\dfrac{i}{2}\Omega_c^{\ast}\hat{s}_{23} \nonumber \\
	&&~~~~~~~~~~~-\dfrac{i}{2}\sigma_{32}\hat{u}_c 
		+\dfrac{i}{2}\sigma_{23}\hat{u}_c^{\dagger} 
		+\hat{F}_{22}, \\
\label{s12}
	&&\dfrac{\partial}{\partial t}\hat{s}_{12} = 
		-(\gamma_{12}-i\delta)\hat{s}_{12} 
		-\dfrac{i}{2}\Omega_p\hat{s}_{32} 
		+\dfrac{i}{2}\Omega_c^{\ast} \hat{s}_{13} \nonumber \\
	&&~~~~~~~~~~~- \dfrac{i}{2}\sigma_{32} \hat{u}_p 
		+\dfrac{i}{2}\sigma_{13}\hat{u}_c^{\dagger} 
		+\hat{F}_{12}, \\
\label{s23}
	&&\dfrac{\partial}{\partial t}\hat{s}_{23} = 
		-\left( \dfrac{\Gamma}{2} - i\Delta_c \right) \hat{s}_{23} 
		+\dfrac{i}{2}\Omega_c(\hat{s}_{22}-\hat{s}_{33}) 
		+\dfrac{i}{2}\Omega_p\hat{s}_{21} \nonumber \\
	&&~~~~~~~~~~~+ \dfrac{i}{2}(\sigma_{22}-\sigma_{33})\hat{u}_c 
		+ \dfrac{i}{2}\sigma_{21}\hat{u}_p 
		+\hat{F}_{23}, \\ 
\label{s13}
	&&\dfrac{\partial}{\partial t}\hat{s}_{13} = 
		-\left( \dfrac{\Gamma}{2} - i\Delta_p \right) \hat{s}_{13} 
		+\dfrac{i}{2}\Omega_p(\hat{s}_{11}-\hat{s}_{33}) 
		+\dfrac{i}{2}\Omega_c\hat{s}_{12} \nonumber \\
	&&~~~~~~~~~~~+ \dfrac{i}{2}(\sigma_{11}
		-\sigma_{33})\hat{u}_p 
		+ \dfrac{i}{2}\sigma_{12}\hat{u}_c 
		+ \hat{F}_{13},\\
\label{s33}
	&&~~~~~~0 = \hat{s}_{11} +\hat{s}_{22} + \hat{s}_{33},
\end{eqnarray}
where $\hat{s}_{ij}$, $\hat{u}_p$ and $\hat{u}_c$ are the fluctuations of $\hat{\sigma}_{ij}$, $\hat{\Omega}_p$ and $\hat{\Omega}_c$, respectively. In the above equations, $\Omega_p(z,t)$, $\Omega_c(z,t)$, and $\sigma_{ij}(z,t)$ are the solutions of the optical Bloch equations and the Maxwell-Schr\"{o}dinger equations of mean fields. Similarly, from Eqs.~(\ref{op}) and (\ref{oc}) we can have the equations for the fluctuation operators of probe and coupling fields as follows. 
\begin{eqnarray}
\label{up}
	&&\left( \dfrac{1}{c}\dfrac{\partial}{\partial t} + \dfrac{\partial}{\partial z} \right) 
		\hat{u}_p =  i \left( \dfrac{\Gamma\alpha}{2L} \right) \hat{s}_{13}, \\
\label{uc}
	&&\left( \dfrac{1}{c}\dfrac{\partial}{\partial t} + \dfrac{\partial}{\partial z} \right) 
		\hat{u}_c =  i \left( \dfrac{\Gamma\alpha}{2L} \right) \hat{s}_{23}.
\end{eqnarray}
When steady-state or continuous-wave cases are considered, all the time derivative terms in Eqs.~(\ref{s11})-(\ref{uc}) are dropped in the calculation.

We focus on the quadrature variance $\langle \Delta\hat{X}^2 \rangle$ of the output probe field, where
\begin{equation}
	\hat{X}(\theta) = e^{-i\theta}\hat{a}_p + e^{i\theta}\hat{a}_p^{\dagger}.
\end{equation}
In the above expression, $\theta$ is the quadrature angle and $\hat{a}_p \equiv \hat{u}_p / g$ (with $g$ being the single-photon Rabi frequency) is the dimensionless fluctuation operator of the probe field. To find $\langle \Delta\hat{X}^2 \rangle$, one needs to know the quantum correlations, i.e., $\langle\hat{a}^{\dagger}_i \hat{a}_j\rangle$, $\langle\hat{a}_i \hat{a}_j\rangle$ where $i,j$ can be $p$ (probe) or $c$ (coupling). In Secs.~3 and 4 of the Appendix, we describe the procedure of converting Eqs.~(\ref{s11})-(\ref{uc}) to the equations of quantum correlations, i.e., Eqs.~(A24)-(A29).
By scanning all quadrature angles, one can find an {\it optimum} quadrature angle, $ \theta_{\rm opt}$, which minimizes the quadrature variance. The variance at $\theta_{\rm opt}$, i.e. degree of squeezing or simply squeezing, is given by
\begin{equation}
\label{Vopt}
	V \equiv \langle \Delta\hat{X}^2 (\theta_{\rm opt}) \rangle  = 
		-\vert\langle\hat{a}_p^2\rangle\vert  -\vert\langle\hat{a}_p^{\dagger 2}\rangle\vert 
		+2\langle\hat{a}_p^{\dagger}\hat{a}_p\rangle +1,
\end{equation}
while $\theta_{\rm opt} = (\text{Arg}[\langle\hat{a}_p^2\rangle] \pm \pi)/2$.
We insert the solutions of Eqs.~(A24)-(A29) into Eq.~(\ref{Vopt}) to determine the output variance.


\section{Optical Density-enhanced squeezed light generation}

It is known that OD of the system ($\alpha$), two-photon detuning ($\delta$), and input Rabi frequencies of the light fields ($\Omega_p$ and $\Omega_c$) are the key factors for the CPT nonlinearity. Consequently, the output squeezings of probe and coupling fields are the functions of these three physical parameters. Since the output squeezings depend significantly on the two-photon detuning of two fields but negligibly on the one-photon detuning  of individual field ($\Delta_p$ or
$\Delta_c$), we consider an asymmetric detuning setting, i.e., $\Delta_p = -\Delta_c = \delta/2$. 
We also set the input probe and coupling Rabi frequencies to be equal, i.e., $\Omega_p(0) = \Omega_c(0) \equiv \Omega$, making the output squeezing of two fields the same. This enables us to report only on the output squeezing of the probe field.

When OD and two-photon detuning are fixed, there exists an optimum input Rabi frequency of light fields to maximize the output squeezing, as demonstrated in Fig.~2(a). The result can be expected by considering the competition between the CPT nonlinearity and light attenuation. A smaller Rabi frequency increases the propagation delay time, i.e., the light-matter interaction time, enhancing nonlinear efficiency to improve the squeezing. On the other hand, a smaller Rabi frequency also causes a larger attenuation of the light under a nonzero two-photon detuning, adding more noises into the system to undermine the squeezing. Hence, a suitable or an optimum input Rabi frequency $\Omega_{\rm opt}$ produces a long interaction time while keeping the attenuation low, resulting in the best squeezing $V_{{\rm opt},\Omega}$ of the output field. 

\FigTwo

Similarly, for a given set of OD and input Rabi frequency, there exists an optimum two-photon detuning to maximize the output squeezing, as demonstrated by Fig.~2(b). At the zero two-photon detuning, the CPT medium becomes completely transparent and there is no nonlinear interaction in the system, resulting in no squeezing at all. A nonzero two-photon detuning introduces the nonlinear interaction and produces the squeezing. However, the two-photon detuning is also accompanied by the attenuation of light, introducing noise to the system. A suitable or an optimum two-photon detuning $\delta_{\rm opt}$ produces a large nonlinearity while keeping the attenuation low, resulting in the best squeezing $V_{{\rm opt},\delta}$ of the output field.

The arguments in the previous two paragraphs, along with the results illustrated in Fig.~2, can also be supported by the analytical solution under some reasonable approximation. Using Eqs.~(\ref{s11})-(\ref{uc}), one can achieve
\begin{equation}
\label{ap}
	\frac{\partial}{\partial \xi} \hat{a}_p = P \hat{a}_p + Q \hat{a}_p^{\dagger} 
		+ R \hat{a}_c + S \hat{a}_c^{\dagger} + \hat{n}_p,
\end{equation}
where $\xi \equiv z/L$ is the dimensionless length, $\hat{a}_c \equiv \hat{u}_c/g$ is similar to the definition of $\hat{a}_p$, $\hat{n}_p = i(\alpha\Gamma/2)\hat{f}_{13}$ where $\hat{f}_{13}$ is the noise operator shown by Eq.~(A14) of the Appendix, and the coefficients $P$, $Q$, $R$, and $S$ are functions of OD $(\alpha)$,  two-photon detuning $(\delta)$, and Rabi frequencies $[\Omega_c(\xi)$ and $\Omega_p(\xi)]$. In the typical CPT experiments, the condition of $\delta\Gamma \ll \Omega^2$ is usually satisfied and the attenuation of probe and coupling fields is comparably  small. We can neglect the attenuation and solve the mean-field equations to obtain 
\begin{equation}
	\Omega_p(\xi) = \Omega_c^*(\xi) \approx \Omega e^{i(\alpha\epsilon/4)\xi}, \nonumber
\end{equation}
where
\begin{equation}
	\epsilon = \dfrac{\Gamma \delta}{\Omega^2}. \nonumber
\end{equation}
With the above $\Omega_p(\xi)$ and $\Omega_c(\xi)$, we find $P$, $Q$, $R$, and $S$ are approximately equal to $\alpha\epsilon^2/8$, $i(\alpha\epsilon/4) \exp[i(\alpha\epsilon/2)\xi]$, $(\alpha\epsilon^2/8) \exp[i(\alpha\epsilon/2)\xi]$, and $\alpha\epsilon^2/8$, respectively.
The magnitudes of $P$, $R$, and $S$ are all small as compared with that of $Q$. Hence, we drop the terms of $P\hat{a}_p$, $R\hat{a}_c$, and $S\hat{a}_c^{\dagger}$ in Eq.~(\ref{ap}), and derive the output variance of the probe field as the following:
\begin{equation}
\label{QplusNoise}
	V \approx \left(\sqrt{|Q|^2 +1} - |Q| \right)^2 + \frac{Z(2+Z)}{3+2Z},
\end{equation}
where
\begin{equation}
	|Q| = \frac{\alpha\epsilon}{4} \nonumber
\end{equation}
is the magnitude of coefficient $Q$, and
\begin{equation}
	Z = \frac{\alpha\epsilon^2}{2} \left(1 + \frac{\Omega^2}{4\Gamma^2} \right) \nonumber
\end{equation}
comes from the noise term $\hat{n}_p$. To achieve Eq.~(\ref{QplusNoise}), we employ not only the conditions of $\epsilon \ll 1$ and $|\Omega_c(\xi)| = |\Omega_p(\xi)| \approx \Omega$, but also that of $|Q|^2 \gg 1$ which is reasonable at a large OD.

In Eq.~(\ref{QplusNoise}), $(\sqrt{|Q|^2 +1} - |Q|)^2$ monotonically decreases with $|Q|$, and $Z(2+Z)/(3+2Z)$ monotonically increases with $Z$. One can immediately see that $|Q|$ produces squeezing and $Z$ deteriorates squeezing. On one hand, $|Q| = \alpha \epsilon/4 = [\alpha\Gamma/(4\Omega^2)] \delta$, and $\alpha\Gamma/(4\Omega^2)$ is about the propagation delay time of light fields in the CPT system. Either a smaller $\Omega$, i.e., a longer delay/interaction time, or a larger $\delta$ makes a larger $|Q|$ and enhances squeezing. On the other hand, as $Z \propto \alpha\epsilon^2/2$, the quantity $\alpha \epsilon^2/2 = \alpha \delta^2\Gamma^2/(2\Omega^4)$ is about the attenuation factor in a CPT system. Either a smaller $\Omega$ or a larger $\delta$, introducing more attenuation of light and adding more noise into the system, makes a larger $Z$ and reduces squeezing. Therefore, there exist optimum $\Omega_{\rm opt}$ and $\delta_{\rm opt}$ to maximize the output squeezing. The analytical expression in Eqs.~(\ref{QplusNoise}) qualitatively explains the behaviors of the numerical results shown in Figs.~2(a) and 2(b).

\FigThree

Since available ODs in experiments are various, we are interested in maximum achievable squeezings at different values of OD. Figures~3(a) and 3(c) illustrate $V_{{\rm opt},\Omega}$ and $\Omega_{\rm opt}$ as functions of the two-photon detuning, where $V_{{\rm opt},\Omega}$ is the maximum squeezing obtained by scanning all input Rabi frequencies. Similarly, Figs.~3(b) and 3(d) show $V_{{\rm opt},\delta}$ and $\delta_{\rm opt}$ as functions of the input Rabi frequency, where $V_{{\rm opt},\delta}$ is the maximum squeezing obtained by scanning all two-photon detunings. We obtained all of the results in Fig.~3 by numerically solving Eqs.~(A24)-(A29). At a given OD, a rather large range of the value of Rabi frequency $\Omega$ or two-photon detuning $\delta$ can achieve similar squeezing as shown by Figs.~3(a) and 3(b). This is expected from Eq.~(\ref{QplusNoise}). Both $|Q|$ and $Z$ depend mainly on $\epsilon$, i.e., $\delta\Gamma/\Omega^2$. Thus, various sets of $\delta$ and $\Omega$ with the same value of $\delta\Gamma/\Omega^2$ all result in similar degrees of squeezing. Figure~3(c) [or Fig.~3(d)] shows the relation between $\Omega_{\rm opt}^2$ and $\delta$ (or between $\sqrt{\delta_{\rm opt}}$ and $\Omega$) forms a nearly straight line, further confirming the above argument. Since the two-photon detuning and input Rabi frequency are easily tunable in experiments, our results imply that the proposed single-passage CPT scheme is very flexible and robust.

A larger OD can always produce smaller variance or larger squeezing as demonstrated by Figs.~3(a) and 3(b). Such result can be understood with the help of Eq.~(\ref{QplusNoise}). To make the variance as small as possible, one must have $|Q|^2 \gg 1$ and $Z \ll 1$. Consequently, $(\sqrt{|Q|^2 +1} - |Q|)^2 \approx 1/|Q|^2 \propto 1/(\alpha^2 \epsilon^2)$ and $Z(2+Z)/(3+2Z) \approx 2Z/3 \propto \alpha \epsilon^2$ in Eq.~(\ref{QplusNoise}). This indicates the optimum $\epsilon \propto \alpha^{-3/4}$. Therefore, the minimum variance is scaled as
\begin{equation}
	V_{\rm opt} \propto \frac{1}{\sqrt{\alpha}},
\end{equation}
which is consistent with the observation on Figs.~3(a) and 3(b), i.e., one-order-of-magnitude increment of OD resulting in 5-dB enhancement of squeezing. Note that since $\epsilon$ is small and the attenuation factor is $\alpha\epsilon^2$ in the CPT system, the probe and coupling transmissions of the data shown in Figs.~3(a) and 3(b) are all larger than 88\%. With an OD of $1,000$, which is accessible by the current technology \cite{OD1,OD2,OD3}, we predict that squeezing of $10$~dB can be achieved. This result demonstrates that the performance of our proposed single-passage CPT scheme is comparable to the state-of-the-art schemes with optical cavities \cite{LIGO,15dB}. 

It is worth to note that the degree of squeezing is affected by relative magnitudes of the probe and coupling Rabi frequencies, $\Omega_p$ and $\Omega_c$. In the CPT case of  $\Omega_p = \Omega_c \equiv \Omega$ discussed here, the squeezing is most prominent. In the EIT case of $\Omega_p \ll \Omega_c$, the squeezing disappears.

\FigFour

We have shown the steady-state quantum fluctuation of output probe field based on the single-passage OD-enhanced CPT scheme. In general, fluctuation is time-dependent. We will discuss the frequency spectrum of output variance under the condition that the squeezing is maximized at the center frequency of probe field. The calculation procedure of spectra can be found in Sec.~V of the Supplemental Material. Figures~4(a) and 4(b) show the spectra of squeezing versus noise frequency ($\omega$) at OD of $1,000$, with two sets of the two-photon detuning ($\delta$) and the input Rabi frequency ($\Omega$). Both sets are optimum and have the same ratio of $\delta$ to $\Omega^2$ that maximizes the squeezing at $\omega = 0$. Similarly, Figs.~4(c) and 4(d) show the spectra at OD of $300$ with two sets of the optimum $\delta$ and $\Omega$. 

The four spectra in Fig.~4 have different bandwidths. At a given OD ($\alpha$), a larger input Rabi frequency (or equivalently a larger two-photon detuning because the ratio of $\delta$ to $\Omega^2$ is fixed) makes the spectrum bandwidth larger. We estimate that the bandwidth approximately follows the formula of $\Omega^2/(\sqrt{2\alpha}\Gamma)$. This completely makes sense, because the width of the CPT transparency window is just proportional to $\Omega^2/(\sqrt{\alpha}\Gamma)$. As for a frequency outside the transparency window, severe attenuation of the light fields adds much noise to destroy the squeezing.

Oscillation behavior is clearly seen in the four spectra of Fig.~4. The comparison between Figs.~4(a) and 4(c) [or between Fig.~4(b) and 4(d)] shows a larger OD makes the oscillation period shorter. In addition, the comparison between Figs.~4(a) and 4(b) [or between Fig.~4(c) and 4(d)] shows a larger input Rabi frequency (or equivalently a larger two-photon detuning) also makes the oscillation period longer. We estimate that the oscillation period roughly follows the formula of $2\pi\times [2\Omega^2/(\alpha\Gamma)]$. In other words, the phase of the oscillation $\phi$ is approximately equal to $[\alpha\Gamma/(2\Omega^2)] \omega$. In the CPT system, the propagation time of light (or light-matter interaction time) $t_d$ is about $\alpha\Gamma/(4\Omega^2)$. Therefore, $\phi \approx 2 \omega t_d$, indicating the light-matter interaction time plays an important role in the oscillation behavior.

Before the conclusion, we want to remark that the ideas to generate squeezed light from EIT systems have been explored~\cite{EITsqueezing1, EITsqueezing2}, the light-atom interaction through an EIT system almost gives a linear response~\cite{sqEIT}. Even though one can apply  the ground-state decoherence to generate squeezed light~\cite{EITsqueezing2}, it is far less efficient than using the two-photon detuning demonstrated in this work. Only currently, the experimental advance of high OD in atomic ensembles has made a great progress, i.e., OD of $1,000$ is readily accessible in several groups. The large OD or long medium length mimics the long effective path length for light in the cavity. Under the same optical path length, slow light can increase the propagation time or interaction time by $5 - 6$ orders of magnitude, while such large dispersive interaction extraordinarily results in a very little attenuation. Utilizing the effect of slow light in the CPT system is another key idea in our proposed scheme

\section{Conclusion}

In summary, through the effect of coherent population trapping (CPT), we have proposed a new concept for the generation of squeezed light from coherent inputs in a single passage. The CPT nonlinearity can be greatly enhanced by the optical density (OD) of the system. An OD of $1,000$, which is accessible by the current technology, produces the squeezing of $10$~dB, and a larger OD can further increase the squeezing. Since the maximum achievable squeezing of a given OD is rather insensitive to the input Rabi frequency or the two-photon detuning individually, both of which are the key parameters in the CPT nonlinearity, the proposed scheme is very flexible and robust. Our study also reveals that the bandwidth in the output squeezing spectra is mainly determined by the width of the CPT transparency window. Combined with light storage and retrieval, squeezed light directly generated from high-OD CPT media has great potentials in the applications of quantum optics and quantum information manipulation utilizing continuous variables.

\section*{Acknowledgment}
The authors acknowledge many fruitful discussions with Dr.\ Julius Ruseckas, in particular on the analytical formula of output variance, under the Taiwan-Latvia-Lithuania cooperation project. This work was supported by the Ministry of Science and Technology of Taiwan under Grant Nos. 105-2628-M-007-003, 105-2119-M-007-004, 105-2923-M-007-002-MY3, and 106-2119-M-007-003. Authors acknowledge many fruitful discussions on this work under the plateforms of TG7 and E1 Programs sponsored by the Physics Division, National Center for Theoretical Sciences, Taiwan.

\appendix
\renewcommand\thesection{A}
\setcounter{equation}{0}
\section*{Appendix} \label{appA}

\subsection{The Heisenberg-Langevin equations}
The derivations to calculate output quadrature variance of fields in the steady-state region are addressed here. First of all, we start with the Heisenberg-Langevin equations for atomic operators $\hat{\sigma}_{\mu\nu}$ from the Hamiltonian given in Eq.~(1) of the main text, i.e.,
\begin{eqnarray}
&&\dfrac{\partial}{\partial t}\hat{\sigma}_{31} = -\left( \dfrac{\Gamma}{2}+i\Delta_p\right)  \hat{\sigma}_{31}-\dfrac{i}{2}(\hat{\sigma}_{11}-\hat{\sigma}_{33})\hat{\Omega}_p^{\dagger} - \dfrac{i}{2}\hat{\Omega}_c^{\dagger}\hat{\sigma}_{21}+\hat{F}_{31}\label{S31},\\
&&\dfrac{\partial}{\partial t}\hat{\sigma}_{32} = -\left( \dfrac{\Gamma}{2}+i\Delta_c\right) \hat{\sigma}_{32}-\dfrac{i}{2}(\hat{\sigma}_{22}-\hat{\sigma}_{33})\hat{\Omega}_c^{\dagger} - \dfrac{i}{2}\hat{\Omega}_p^{\dagger}\hat{\sigma}_{12}+\hat{F}_{32}\label{S32},\\
&&\dfrac{\partial}{\partial t}\hat{\sigma}_{21} = -\left( {\gamma}_{12}+i\delta\right) \hat{\sigma}_{21}+\dfrac{i}{2}\hat{\Omega}_p^{\dagger}\hat{\sigma}_{23} - \dfrac{i}{2}\hat{\sigma}_{31}\hat{\Omega}_c+\hat{F}_{21}\label{S21},\\
&&\dfrac{\partial}{\partial t}\hat{\sigma}_{11} = \dfrac{\Gamma}{2} \hat{\sigma}_{33}-\dfrac{i}{2}\hat{\sigma}_{31}\hat{\Omega}_p + \dfrac{i}{2} \hat{\Omega}_p^{\dagger}\hat{\sigma}_{13}+\hat{F}_{11}\label{S11},\\
&&\dfrac{\partial}{\partial t}\hat{\sigma}_{22} = \dfrac{\Gamma}{2} \hat{\sigma}_{33}-\dfrac{i}{2}\hat{\sigma}_{32}\hat{\Omega}_c + \dfrac{i}{2} \hat{\Omega}_c^{\dagger}\hat{\sigma}_{23}+\hat{F}_{22}\label{S22},\\
&&\dfrac{\partial}{\partial t}\hat{\sigma}_{33} = -\Gamma \hat{\sigma}_{33}+\dfrac{i}{2}\hat{\sigma}_{31}\hat{\Omega}_p+\dfrac{i}{2}\hat{\sigma}_{32}\hat{\Omega}_c -\dfrac{i}{2} \hat{\Omega}_p^{\dagger}\hat{\sigma}_{13}-\dfrac{i}{2} \hat{\Omega}_c^{\dagger}\hat{\sigma}_{23}+\hat{F}_{33}\label{S33},\\
&&\dfrac{\partial}{\partial t}\hat{\sigma}_{12} = -\left( {\gamma}_{12}-i\delta\right) \hat{\sigma}_{12}-\dfrac{i}{2}\hat{\sigma}_{32}\hat{\Omega}_p + \dfrac{i}{2}\hat{\Omega}_c^{\dagger}\hat{\sigma}_{13}+\hat{F}_{12}\label{S12},\\
&&\dfrac{\partial}{\partial t}\hat{\sigma}_{23} = -\left( \dfrac{\Gamma}{2}-i\Delta_c\right) \hat{\sigma}_{23}+\dfrac{i}{2}(\hat{\sigma}_{22}-\hat{\sigma}_{33})\hat{\Omega}_c + \dfrac{i}{2}\hat{\sigma}_{21}\hat{\Omega}_p+\hat{F}_{23}\label{S23},\\
&&\dfrac{\partial}{\partial t}\hat{\sigma}_{13} = -\left( \dfrac{\Gamma}{2}-i\Delta_p\right) \hat{\sigma}_{13}+\dfrac{i}{2}(\hat{\sigma}_{11}-\hat{\sigma}_{33})\hat{\Omega}_p + \dfrac{i}{2}\hat{\sigma}_{12}\hat{\Omega}_c+\hat{F}_{13}\label{S13}.
\end{eqnarray}
For steady-state case, we drop the time derivation in the left hand side. Then, we find the mean-field solutions both for field and atomic operators, and expand the product of any two operators to first-order of quantum fluctuations, i.e. $\hat{A}\hat{B}\simeq AB + A\hat{b} + B\hat{a}$. Here, $A (B)$ and $\hat{a} (\hat{b})$ denote the mean-field and corresponding quantum fluctuation of operator $ \hat{A}(\hat{B})$.  As the terms related to $\hat{a}\hat{b}$ are much smaller, then one can safely ignore them.
  
As for the propagation equations for optical fields, shown in Eqs. (4) and (5) of the main text, we also separate the corresponding mean-field and their quantum fluctuation by the same procedure. 
When keeping the nonlinear effects in the mean-field equations,  we can obtain a set of linearized  equations of motion for quantum fluctuations. In the following, we show the process in details to obtain and solve mean-fields and quantum fluctuations with a systematic approach. 

\subsection{Mean-Field Solutions}
To have a clear illustration, we rewrite the mean-field part of Eqs.~(\ref{S31})-(\ref{S13}) into a matrix form, i.e., $\textbf{M}_\textbf{1} \textbf{x} = \textbf{b}$, with  $\textbf{M}_1$ explicitly written as
\begin{eqnarray}
\textbf{M}_\textbf{1} = \left( \begin{array}{ccccccccc} 
-\tilde{\gamma}_{13}^{\ast}  & 0 & -i\dfrac{\Omega_c^{\ast}}{2} & -i\dfrac{\Omega_p^{\ast}}{2} & 0 & i\dfrac{\Omega_p^{\ast}}{2} & 0 & 0 & 0 \\
0 & -\tilde{\gamma}_{23}^{\ast} & 0 & 0 & -i\dfrac{\Omega_c^{\ast}}{2} & i\dfrac{\Omega_c^{\ast}}{2} & -i\dfrac{\Omega_p^{\ast}}{2} & 0 & 0\\
-i\dfrac{\Omega_c}{2} & 0 & -\left( \gamma_{12}+i\delta\right) & 0 & 0 & 0 & 0 & i\dfrac{\Omega_p^{\ast}}{2} & 0\\
-i\dfrac{\Omega_p}{2} & 0 & 0 & 0 & 0 & \Gamma/2 & 0 & 0 & i\dfrac{\Omega_p^{\ast}}{2} \\
0 & -i\dfrac{\Omega_c}{2} & 0 & 0 & 0 & \Gamma/2 & 0 & i\dfrac{\Omega_c^{\ast}}{2} & 0 \\
0 & 0 & 0 & 1 & 1 & 1 & 0 & 0 & 0 \\
0 & -i\dfrac{\Omega_p}{2} & 0 & 0 & 0 & 0 & -\left( \gamma_{12}-i\delta\right) & 0 & i\dfrac{\Omega_c^{\ast}}{2}\\
0 & 0 & i\dfrac{\Omega_p}{2} & 0 & i\dfrac{\Omega_c}{2} & -i\dfrac{\Omega_c}{2} & 0 & -\tilde{\gamma}_{23} & 0\\
0 & 0 & 0 & i\dfrac{\Omega_p}{2} & 0 & -i\dfrac{\Omega_p}{2} & i\dfrac{\Omega_c}{2} & 0 & -\tilde{\gamma}_{13}
\end{array}\right)_{9\times 9},
\label{Mmatrix}
\end{eqnarray}
where $ \tilde{\gamma}_{13} \equiv  \Gamma/2-i\Delta_p $ and $ \tilde{\gamma}_{23} \equiv  \Gamma/2-i\Delta_c $.  Here, the notations are defined as $ \textbf{x}^{T} = (\sigma_{31},\sigma_{32},\sigma_{21},\sigma_{11},\sigma_{22},\sigma_{33},\sigma_{12},\sigma_{23},\sigma_{13} )$ and 
$ \textbf{b}^T = (0, 0, 0, 0, 0, 1, 0, 0, 0) $.
In Eq.~(\ref{Mmatrix}), we also have replaced Eq.~(\ref{S33}) with the help of population conservation, i.e., $ \sigma_{11}+\sigma_{22}+\sigma_{33} = 1$.

The corresponding steady-state solution can be easily obtained by using the matrix algebra: $\textbf{x} = \textbf{M}_\textbf{1}^{-1}\textbf{b}$.
The two dipole sources $\sigma_{13} = \sigma_{13}(\Omega_p,\Omega_p^{\ast},\Omega_c, \Omega_c^{\ast})$ and $\sigma_{23} = \sigma_{23}(\Omega_p,\Omega_p^{\ast},\Omega_c, \Omega_c^{\ast})$ count the nonlinear responses with respect to the optical fields.
With the solutions of $\sigma_{13}$ and $\sigma_{23}$ substituted into the mean-field part of Eqs.~(4) and (5) of the main text, one can obtain the corresponding solutions for optical fields.

\subsection{Quantum Fluctuation Solutions}

For the fluctuation operators in the atomic parts, as well as their hermitian conjugates, we linearize Eqs.~(\ref{S31})-(\ref{S13}) to obtain the results shown in Eqs.~(6)-(10) of the main text.
Again, in the steady-state, we express the fluctuation operators in a matrix form: $\textbf{M}_\textbf{1}~\textbf{y} + \textbf{M}_\textbf{2}~\textbf{u} + \textbf{r} = 0$, where $ \textbf{y}^{T} = \left( \hat{s}_{31},  \hat{s}_{32},  \hat{s}_{21},  \hat{s}_{11},  \hat{s}_{22}, \hat{s}_{33}, \hat{s}_{12}, \hat{s}_{23}, \hat{s}_{13} \right)$ gives  the fluctuations of atomic operators,  $\textbf{u}^{T} = \left( \hat{u}_p,  \hat{u}_p^{\dagger}, \hat{u}_c,  \hat{u}_c^{\dagger} \right)$ denotes the fluctuations of field operators, and $\textbf{r}^{T} = \left( \hat{F}_{31},  \hat{F}_{32},  \hat{F}_{21},  \hat{F}_{11},  \hat{F}_{22}, \hat{F}_{33}, \hat{F}_{12}, \hat{F}_{23}, \hat{F}_{13} \right)$ are the corresponding Langevin noise operators, respectively. The matrix $ \textbf{M}_\textbf{2} $ is a 9 by 4 matrix, with the matrix elements having the form:
\begin{eqnarray}
\textbf{M}_\textbf{2} = \dfrac{1}{2}\left( \begin{array}{cccc} 
0 & -i\left( \sigma_{11}-\sigma_{33}\right) & 0 & -i\sigma_{21} \\
0 & -i\sigma_{12} & 0 & -i\left( \sigma_{22}-\sigma_{33}\right) \\
0 & i\sigma_{23} & -i\sigma_{31} & 0 \\
-i\sigma_{31} & i\sigma_{13} & 0 & 0 \\
0 & 0 & -i\sigma_{32} & i\sigma_{23} \\
0 & 0 & 0 & 0 \\
-i\sigma_{32} & 0 & 0 & i\sigma_{13} \\
i\sigma_{21} & 0 & i\left( \sigma_{22}-\sigma_{33}\right) & 0 \\
i\left( \sigma_{11}-\sigma_{33}\right) & 0 & i\sigma_{12} & 0
\end{array}\right)_{9\times 4}. \label{Mmatrix2}
\end{eqnarray}
The atomic fluctuation part can be found by solving $\textbf{y} = \textbf{T}\textbf{M}_\textbf{2}~\textbf{u} + \textbf{T}\textbf{r}$,  where we define $\textbf{T} \equiv -\textbf{M}_{\textbf{1}}^{-1}$. 
With the solution of $ \textbf{y} $, we can have the expressions for the quantum fluctuations in  two dipole sources $\hat{s}_{13} = \textbf{y}(9) $ and $\hat{s}_{23} = \textbf{y}(8)$,  in terms of the field fluctuation operators. 
In general, one can write down $\hat{s}_{13}$ and $\hat{s}_{23}$ in the following form:
\begin{eqnarray}
\hat{s}_{13} = A_1 \hat{u}_p + B_1 \hat{u}_p^{\dagger} + C_1 \hat{u}_c + D_1 \hat{u}_c^{\dagger} + \hat{f}_{13}\label{s13},\\
\hat{s}_{23} = A_2 \hat{u}_p + B_2 \hat{u}_p^{\dagger} + C_2 \hat{u}_c + D_2 \hat{u}_c^{\dagger} + \hat{f}_{23}\label{s23}.
\end{eqnarray}
Here, $A_i,~B_i,~C_i $ and $ D_i $ can be directly calculated from the matrix  $\textbf{T}\textbf{M}_{\textbf{2}}$, with the  effective Langevin noise operators $\hat{f}_{13}$ and $\hat{f}_{23}$ obtained from the 9th and 8th elements of $\textbf{T}\textbf{r}$.
Moreover, we also have $\hat{f}^{\dagger}_{ij} = \hat{f}_{ji}$.
In particular, the explicit forms for $\hat{f}_{13}$ and $\hat{f}_{23}$ can be found to be
\begin{eqnarray}
\begin{split}
&\hat{f}_{13} = \left( \textbf{T}\textbf{r}\right)_9 = \sum_{k =1}^9 T_{9k}~r_k, \\
&\hat{f}_{23} = \left( \textbf{T}\textbf{r}\right)_8 = \sum_{k =1}^9 T_{8k}~r_k .
\end{split}
\label{fTr}
\end{eqnarray}

At the same time, we can obtain the steady-state solutions for fields from the propagation equation shown in  Eqs.~(12) and (13) of the main text. They are
\begin{eqnarray}
\label{fpropagation1}
\dfrac{\partial}{\partial\xi}\hat{u}_p &=& i\left( \dfrac{\Gamma\alpha}{2}\right)\hat{s}_{13},\\
\label{fpropagation2}
\dfrac{\partial}{\partial\xi}\hat{u}_c &=& i\left( \dfrac{\Gamma\alpha}{2}\right)\hat{s}_{23},
\end{eqnarray}
with a dimensionless length denoted as $\xi \equiv z/L$.

By substituting Eqs.~(\ref{s13}) and (\ref{s23}) and their hermitian conjugates into the propagation equation for field fluctuations shown in Eqs.~(\ref{fpropagation1}) and (\ref{fpropagation2}), we can obtain a compact form for the noise operators for fields $\textbf{a}^{\textbf{T}} \equiv \left( \hat{a}_p, \hat{a}_p^{\dagger}, \hat{a}_c, \hat{a}_c^{\dagger}\right)$:
\begin{eqnarray}
\dfrac{\partial}{\partial\xi}\textbf{a} = \textbf{C}~\textbf{a} + \textbf{N}.
\label{axi}
\end{eqnarray} 
Here,  the two matrices of $\textbf{C} $ and $\textbf{N} $ have the explicit form as
\begin{eqnarray}
&&\textbf{C} = i\dfrac{\Gamma \alpha}{2}\left( \begin{array}{cccc}
A_1 & B_1 & C_1 & D_1 \\
-B_1^{\ast} & -A_1^{\ast} & -D_1^{\ast} & -C_1^{\ast} \\
A_2 & B_2 & C_2 & D_2 \\
-B_2^{\ast} & -A_2^{\ast} & -D_2^{\ast} & -C_2^{\ast} 
\end{array}\right)\equiv
\left( \begin{array}{cccc}
P_1 & Q_1 & R_1 & S_1 \\
Q_1^{\ast} & P_1^{\ast} & S_1^{\ast} & R_1^{\ast} \\
P_2 & Q_2 & R_2 & S_2 \\
Q_2^{\ast} & P_2^{\ast} & S_2^{\ast} & R_2^{\ast} 
\end{array}\right),\label{C}\\
&&\textbf{N} = i\dfrac{\Gamma \alpha}{2~g}\left( \begin{array}{c}
\hat{f}_{13},
-\hat{f}_{13}^{\dagger},
\hat{f}_{23},
-\hat{f}_{23}^{\dagger}
\end{array}\right)^{T}.
\label{N}
\end{eqnarray}

\subsection{Equations of Motion for Quantum Correlations}

In order to calculate the quadrature variance in the output fields, we have to know the corresponding field-field correlations. 
According to Eqs.~(\ref{axi})-(\ref{N}), one can obtain the equations of motion for all the two-field correlations in the following form
\begin{eqnarray}
\dfrac{\partial}{\partial\xi}\langle\textbf{a}\textbf{a}^{\dagger}\rangle = \textbf{C}~\langle\textbf{a}\textbf{a}^{\dagger}\rangle + \langle\textbf{a}\textbf{a}^{\dagger}\rangle~\textbf{C}^{\dagger} + \textbf{Z}.
\label{aad}
\end{eqnarray}
Here, the matrix $\textbf{Z}$ shows the correlations of Langevin noise operators, denoted  $\langle \textbf{N}\textbf{N}^{\dagger}\rangle$. That is 
\begin{eqnarray}
\textbf{Z}\equiv\langle \textbf{N}\textbf{N}^{\dagger}\rangle = \dfrac{\Gamma \alpha}{4} \left( \textbf{V} \mathcal{D} \textbf{V}^{\dagger}\right).
\label{NN}
\end{eqnarray}
Here, we have applied the matrix product of $ \textbf{T}\textbf{r} $ and the correlations of any two Langevin noise operators, i.e., $\langle\hat{F}_{\mu}\hat{F}_{\nu}\rangle = \mathcal{D}_{\mu\nu} \, c/(NL)$.
The diffusion coefficient,  $\mathcal{D}_{\mu\nu}$,  can be obtained from the generalized Einstein relation \cite{diffusionD}.
Moreover, to link the optical density (OD) and the related single photon Rabi frequency, we also define $\alpha = g^2 N L/ (c\Gamma)$. The matrix \textbf{V} shown in Eq.~(\ref{NN}) has the form:
\begin{eqnarray}
\textbf{V} \equiv \left( \begin{array}{ccccccccc}
T_{91} & T_{92} & T_{93} & T_{94} & T_{95} & T_{96} & T_{97} & T_{98} &  T_{99} \\
-T_{11} & -T_{12} & -T_{13} & -T_{14} & -T_{15} & -T_{16} & -T_{17} & -T_{18} & -T_{19} \\
T_{81} & T_{82} & T_{83} & T_{84} & T_{85} & T_{86} & T_{87} & T_{88} &  T_{89} \\
-T_{21} & -T_{22} & -T_{23} & -T_{24} & -T_{25} & -T_{26} & -T_{27} & -T_{28} &  -T_{29} 
\end{array}\right)_{4\times 9},
\label{V}
\end{eqnarray}
with the corresponding diffusion coefficeints in the matrix $\mathcal{D}$:
\begin{eqnarray}
\mathcal{D} = \left( \begin{array}{ccccccccc}
0 & 0 & 0 & 0 & 0 & 0 & 0 & 0 & 0 \\
0 & 0 & 0 & 0 & 0 & 0 & 0 & 0 & 0 \\
0 & 0 & \gamma_2\sigma_{33} & 0 & 0 & 0 & 0 & 0 & 0 \\
0 & 0 & 0 & \gamma_1\sigma_{33} & 0 & 0 & 0 & -\gamma_1\sigma_{32} & -\gamma_1\sigma_{31} \\
0 & 0 & 0 & 0 & \gamma_2\sigma_{33} & 0 & 0 & -\gamma_2\sigma_{32} & -\gamma_2\sigma_{31} \\
0 & 0 & 0 & 0 & 0 & 0 & 0 & 0 & 0 \\
0 & 0 & 0 & 0 & 0 & 0 & \gamma_1\sigma_{33} & 0 & 0 \\
0 & 0 & 0 & -\gamma_1\sigma_{23} & -\gamma_2\sigma_{23} & 0 & 0 & \gamma_2\sigma_{33}+\Gamma\sigma_{22} & \Gamma\sigma_{21} \\
0 & 0 & 0 & -\gamma_1\sigma_{13} & -\gamma_2\sigma_{13} & 0 & 0 & \Gamma\sigma_{12} & \gamma_1\sigma_{33}+\Gamma\sigma_{11} 
\end{array}
\right)_{9\times 9} .
\label{rrd}
\end{eqnarray}
Based on Eq.~(\ref{aad}), and with the help of Eqs.~(\ref{C}) and (\ref{N}) and Eqs.~(\ref{NN})-(\ref{rrd}), the equations of motion for quantum correlations can be found as:
\begin{eqnarray}
&&\dfrac{\partial}{\partial \xi}\langle\hat{a}_p^2\rangle = 2P_1\langle\hat{a}_p^2\rangle + Q_1\left( 2\langle\hat{a}_p^{\dagger}\hat{a}_p\rangle + 1\right) + 2 R_1 \langle\hat{a}_p\hat{a}_c\rangle + 2S_1\langle\hat{a}_p\hat{a}_c^{\dagger}\rangle + n_1 \label{corr1}\\
&&\dfrac{\partial}{\partial \xi}\langle\hat{a}_p^{\dagger}\hat{a}_p\rangle = 2P_1'\langle\hat{a}_p^{\dagger}\hat{a}_p\rangle + Q_1^{\ast}\langle\hat{a}_p^2\rangle + Q_1 \langle\hat{a}_p^{\dagger 2}\rangle + S_1^{\ast}\langle\hat{a}_p\hat{a}_c\rangle + S_1\langle\hat{a}_p^{\dagger}\hat{a}_c^{\dagger}\rangle + R_1^{\ast}\langle\hat{a}_p\hat{a}_c^{\dagger}\rangle + R_1\langle\hat{a}_p^{\dagger}\hat{a}_c\rangle + n_2 \label{corr2}\\
&&\dfrac{\partial}{\partial \xi}\langle\hat{a}_c^2\rangle = 2R_2\langle\hat{a}_c^2\rangle + 2 P_2 \langle\hat{a}_p\hat{a}_c\rangle + 2Q_2\langle\hat{a}_p^{\dagger}\hat{a}_c\rangle + S_2\left( 2\langle\hat{a}_c^{\dagger}\hat{a}_c\rangle + 1\right) + n_3 \label{corr3}\\
&&\dfrac{\partial}{\partial \xi}\langle\hat{a}_c^{\dagger}\hat{a}_c\rangle =  2R_2'\langle\hat{a}_c^{\dagger}\hat{a}_c\rangle  + Q_2^{\ast}\langle\hat{a}_p\hat{a}_c\rangle + Q_2\langle\hat{a}_p^{\dagger}\hat{a}_c^{\dagger}\rangle + P_2^{\ast}\langle\hat{a}_p^{\dagger}\hat{a}_c\rangle + P_2\langle\hat{a}_p\hat{a}_c^{\dagger}\rangle + S_2^{\ast}\langle\hat{a}_c^2\rangle + S_2 \langle\hat{a}_c^{\dagger 2}\rangle + n_4 \label{corr4}\\
&&\dfrac{\partial}{\partial \xi}\langle\hat{a}_p\hat{a}_c\rangle = \left( P_1+R_2\right) \langle\hat{a}_p\hat{a}_c\rangle + R_1\langle\hat{a}_c^2\rangle + P_2 \langle\hat{a}_p^2\rangle + S_1 \langle\hat{a}_c^{\dagger}\hat{a}_c\rangle + S_2 \langle\hat{a}_p\hat{a}_c^{\dagger}\rangle + Q_1\langle\hat{a}_p^{\dagger}\hat{a}_c\rangle + Q_2\left( \langle\hat{a}_p^{\dagger}\hat{a}_p\rangle + 1\right) + n_5 ~~~~~\label{corr5}\\
&&\dfrac{\partial}{\partial \xi}\langle\hat{a}_p^{\dagger}\hat{a}_c\rangle = \left( P_1^{\ast}+R_2\right) \langle\hat{a}_p^{\dagger}\hat{a}_c\rangle + Q_1^{\ast}\langle\hat{a}_p\hat{a}_c\rangle + Q_2 \langle\hat{a}_p^{\dagger 2}\rangle + S_1^{\ast}\langle\hat{a}_c^2\rangle + R_1^{\ast}\langle\hat{a}_c^{\dagger}\hat{a}_c\rangle + P_2 \langle\hat{a}_p^{\dagger}\hat{a}_p\rangle + S_2 \langle\hat{a}_p^{\dagger}\hat{a}_c^{\dagger}\rangle + n_6 \label{corr6},
\end{eqnarray} 
here, $ P_1' $ and $ R_2' $ denote the real parts of $ P_1 $ and $ R_2 $; while $n_i~(i = 1-6) $ are the corresponding noise-noise correlations.
The explicit expressions for $ n_i $ have the forms:
\begin{eqnarray}
&&n_1 = \textbf{Z}(1,2) = -\eta \langle\hat{f}_{13}\hat{f}_{13}\rangle,\\
&&n_2 = \textbf{Z}(2,2) = +\eta \langle\hat{f}_{13}^{\dagger}\hat{f}_{13},\rangle\\
&&n_3 = \textbf{Z}(3,4) = -\eta \langle\hat{f}_{23}\hat{f}_{23}\rangle,\\
&&n_4 = \textbf{Z}(4,4) = +\eta \langle\hat{f}_{23}^{\dagger}\hat{f}_{23}\rangle,\\
&&n_5 = \textbf{Z}(1,4) = -\eta \langle\hat{f}_{13}\hat{f}_{23}\rangle,\\
&&n_6 = \textbf{Z}(2,4) = +\eta \langle\hat{f}_{13}^{\dagger}\hat{f}_{23}\rangle,
\end{eqnarray}
with $ \eta \equiv [\Gamma \alpha/(2g)]^2 $.
As the case of coherent state inputs is considered, the initial conditions at $\xi = 0$ for these correlations given in Eqs.~(\ref{corr1})-(\ref{corr6}) are set to be zeros. 
By solving Eqs.~(\ref{corr1})-(\ref{corr6}) directly, one can find the corresponding minimum value in the quadrature variance, as shown in Eq.~(15) of the main text. 

\begin{figure}[t]
\centering
\includegraphics[scale=0.5]{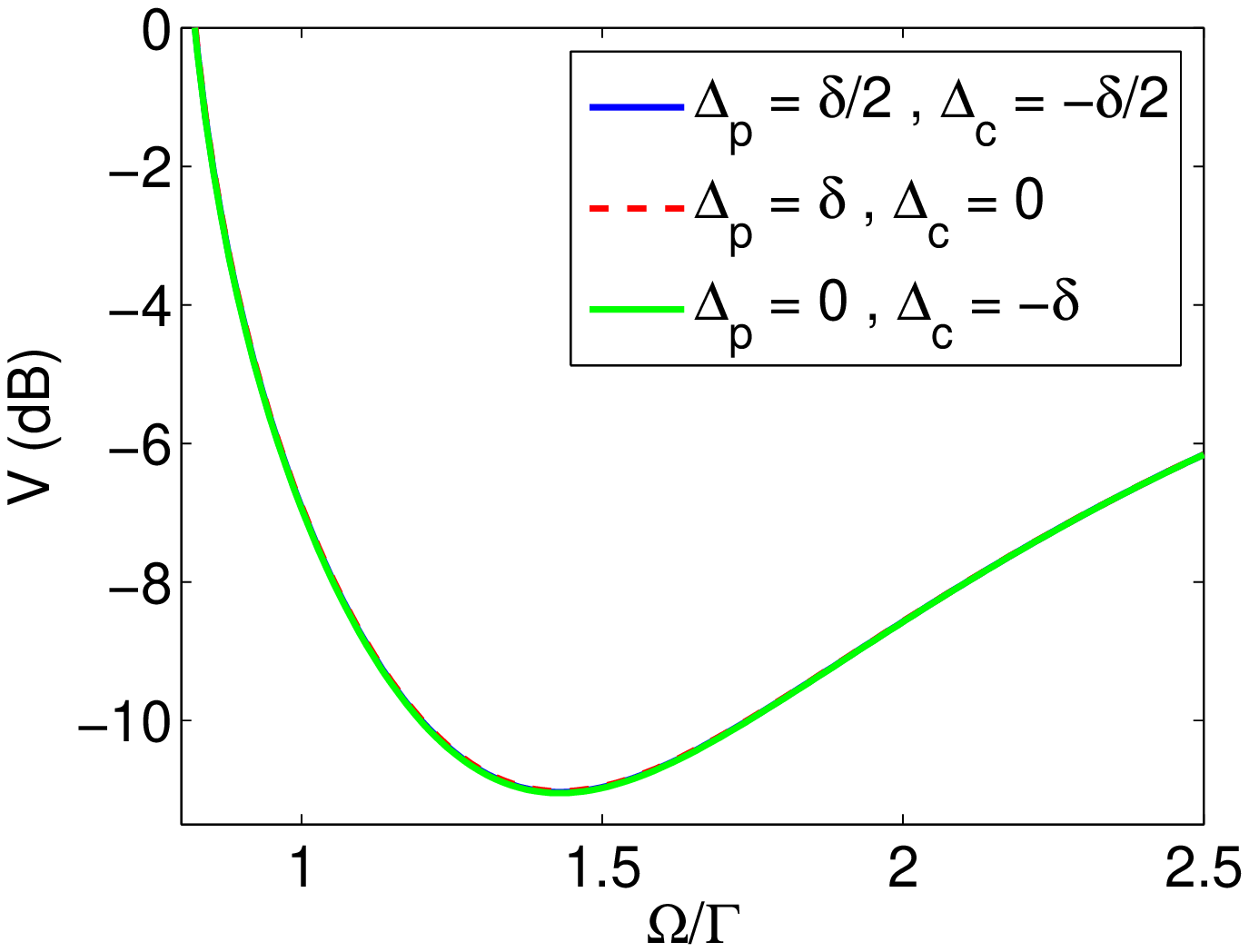} 
\includegraphics[scale=0.5]{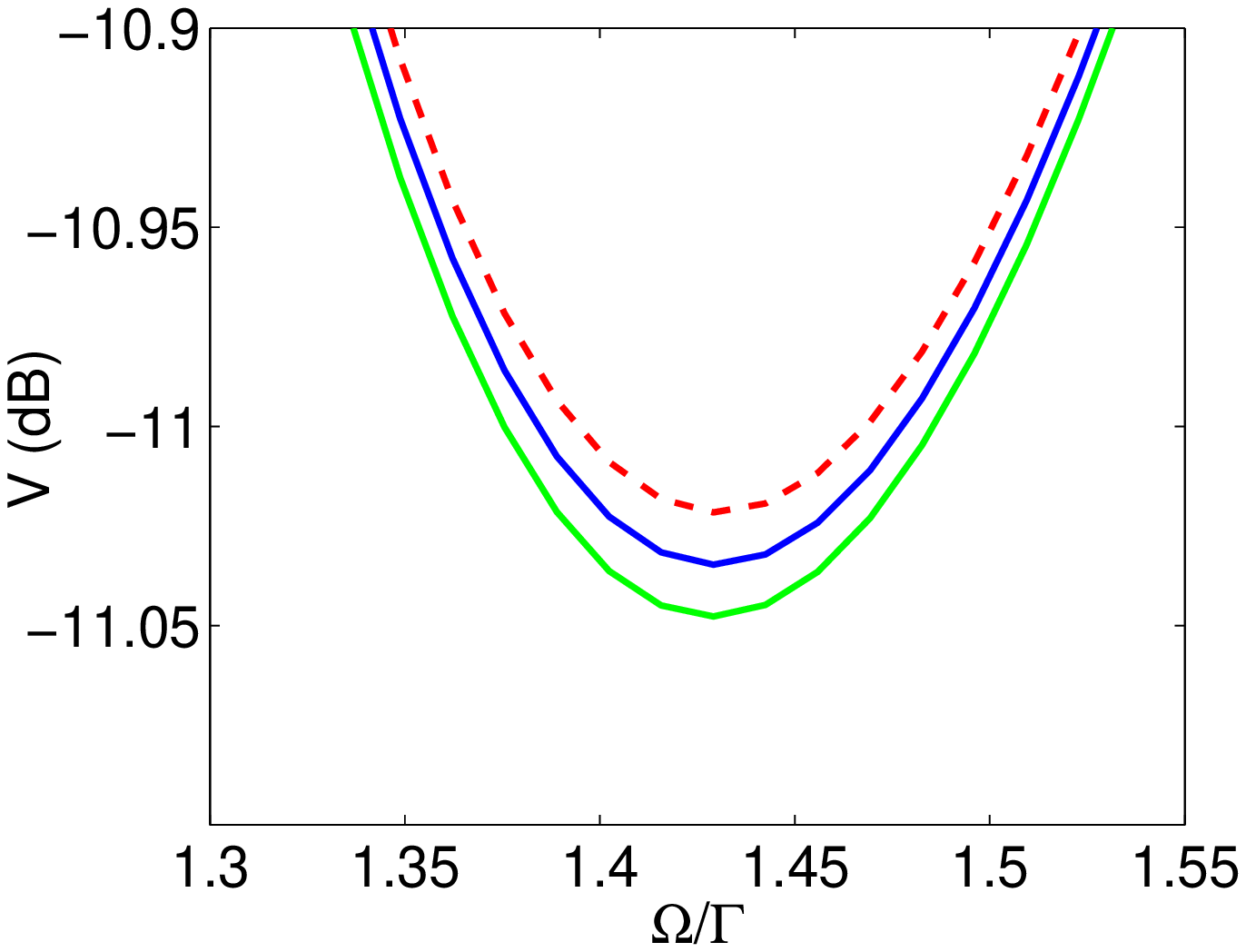} 
\caption{Left panel: The optimized (minimum)  output variances v.s. the normalized Rabi frequency $\Omega/\Gamma$ with three different detuning settings. Here, we have set $\Omega_p = \Omega_c = 1\Gamma$ and $\delta = \pm 0.02\Gamma$. Right panel: Same as the Left panel, but with an enlarged plot around the  the minimum variance,  for recognizing the difference between these three detuning settings. 
\label{s1}}
\end{figure}

\subsection{Asymmetric Detuning and Unequal Rabi Frequencies of Two Optical Fields}

In this Section, we discuss the arrangement of asymmetric detuning and unequal  Rabi frequencies of two optical fields on the squeezed light generation. 
For the asymmetric detuning settings, we choose $ \Delta_p = -\Delta_c = \delta/2 $ as an illustration.  By keeping the value of two-photon detuning as a constant, i.e., $\Delta_p - \Delta_c = \delta$, in  Fig.~\ref{s1}, we reveal the optimized (minimum) output variance $V$ as a function of normalized Rabi frequency $\Omega/\Gamma$.

\begin{figure}[h]
\centering
\includegraphics[scale=0.5]{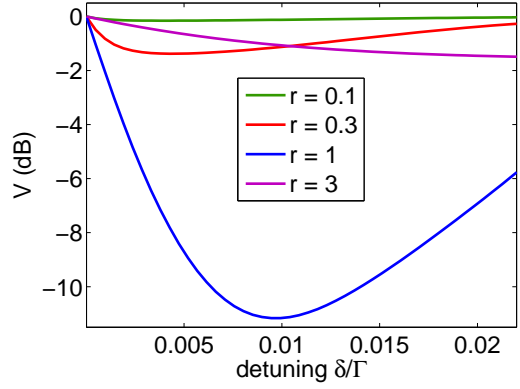} 
\caption{The output variances v.s. the normalized two-photon detuning $\delta/\Gamma$ with four different Rabi frequency ratios $r$. Here, we have considered $\Delta_p = -\Delta_c = \delta/2$ and $\Omega_c = 1\Gamma$.
\label{s2}}
\end{figure}

From Fig. \ref{s1}, one can see only a slight difference in these three detuning settings. 
It means that in our proposed scheme the output variance is almost insensitive to the detuning setting. 
Similarly, one can also consider the setting with unequal Rabi frequencies of two optical fields. 
To give a clear comparison,  we denote the ratio between probe and coupling Rabi frequencies as $r \equiv \Omega_p/\Omega_c$.
Then, in Fig. \ref{s2}, we show the resulting output variance as a function of two-photon detuning $\delta/\Gamma$ for different ratios. 
It is clear to see that we have the strongest squeezing in the output variance when $r = 1$, which just discloses the condition of CPT.
Meanwhile, the squeezing effect becomes very vague when the ratio is much larger or smaller than $1$. 
For example, in the setting with $r = 0.1$,  which corresponds to the condition of EIT,  no significant squeezing can be seen as the curve in Green-color shown in Fig. \ref{s2}.

\subsection{Squeezing Spectra in the Output Fields}

The quadrature variance in the output fields can be measured directly in experiments. 
To calculate the variance spectrum for the output fields, we need to take the time-dependent fluctuations of field and atomic operators into account.   
Here, we perform the  Fourier transform for all the fluctuation operators into the frequency domain, i.e., $\hat{O}(t) \rightarrow \tilde{O}(\omega)$. 
For the atomic fluctuations, we have
\begin{eqnarray}
\tilde{\textbf{y}} = \textbf{T}' \textbf{M}_2 \textbf{u} + \textbf{T}'\textbf{r},
\end{eqnarray}
with $\tilde{\textbf{y}}^{T} = \left[ \tilde{s}_{31}(\omega),  \tilde{s}_{32}(\omega),  \tilde{s}_{21}(\omega),  \tilde{s}_{11}(\omega),  \tilde{s}_{22}(\omega), \tilde{s}_{33}(\omega), \tilde{s}_{12}(\omega), \tilde{s}_{23}(\omega), \tilde{s}_{13}(\omega) \right]$, and $\textbf{T}' = -\left( \textbf{M}_1 + i\omega \textbf{I}_o\right) ^{-1}$.
Here,  $ \textbf{I}_o $ is a matrix whose non-zero matrix elements are $1$ in the diagonal part,  but only with $ \textbf{I}_o(6,6) = 0$.
In the frequency domain , the propagation equations for field fluctuations are given by
\begin{eqnarray}
&&\dfrac{\partial}{\partial\xi}\tilde{a}_p(\omega) = i\dfrac{\omega L}{c}\tilde{a}_p(\omega) + i\dfrac{\Gamma \alpha}{2}\left[ A_1'(\omega)\tilde{a}_p(\omega) + B_1'(\omega) \tilde{a}_p^{\dagger}(-\omega) + C_1'(\omega)\tilde{a}_c(\omega) + D_1'(\omega) \tilde{a}_c^{\dagger}(-\omega) + \dfrac{\tilde{f}_{13}(\omega)}{g}\right],\label{apw}\\
&&\dfrac{\partial}{\partial\xi}\tilde{a}_c(\omega) = i\dfrac{\omega L}{c}\tilde{a}_c(\omega) + i\dfrac{\Gamma \alpha}{2}\left[ A_2'(\omega)\tilde{a}_p(\omega) + B_2'(\omega) \tilde{a}_p^{\dagger}(-\omega) + C_2'(\omega)\tilde{a}_c(\omega) + D_2'(\omega) \tilde{a}_c^{\dagger}(-\omega) + \dfrac{\tilde{f}_{23}(\omega)}{g}\right], \label{acw}
\end{eqnarray}
where we have the coefficients: $ A_{1,2}'(\omega) $, $ B_{1,2}'(\omega) $, $ C_{1,2}'(\omega) $, and $ D_{1,2}'(\omega)$, obtained from the matrix product of $ \textbf{T}' \textbf{M}_2$ accordingly. 
As the quadrature operator in the output probe field is defined as
$\tilde{X}_p(\omega) \equiv \tilde{a}_p(\omega) + \tilde{a}_p^{\dagger}(-\omega)$, we can calculate the optima squeezing spectrum through the following formula:
\begin{eqnarray}
S(\omega) \equiv \langle\tilde{X}(\omega)\tilde{X}^{\dagger}(\omega)\rangle
= -\vert\langle\tilde{a}_p(\omega)\tilde{a}_p(-\omega)\rangle\vert
-\vert\langle\tilde{a}_p^{\dagger}(\omega)\tilde{a}_p^{\dagger}(-\omega)\rangle\vert + \langle\tilde{a}_p^{\dagger}(-\omega)\tilde{a}_p(-\omega)\rangle + \langle\tilde{a}_p(\omega)\tilde{a}_p^{\dagger}(\omega)\rangle. \label{SW}
\end{eqnarray}
With Eqs.~(\ref{apw})-(\ref{SW}), the squeezing spectrum shown in Fig.~4 of the main text can be generated.



\begin{thebibliography}{1}
\bibitem{gamma}
A non-negligible ground-state decoherence rate $\gamma_{12}$ can also produce the squeezing of output fields in CPT media. However, inducing the squeezing by $\gamma_{12}$, which is beyond the scope of this work, is less effective than by the two-photon detuning.
\end{thebibliography}
\end{document}